\documentclass[useAMS,usenatbib,psfig]{mn2e}
\usepackage{psfig}
\usepackage{times}

\begin{document}

\title[Effect of LSS on the Galaxy 3--Point Function]{The Effect of
  Large-Scale Structure on the SDSS Galaxy Three--Point Correlation
  Function} 
\author[Nichol, R. C., et al.]{ \parbox{\textwidth}{ R. C.
    Nichol$^{1,2}$, Ravi K. Sheth$^{3}$, Y. Suto$^{4}$, A. J. Gray$^{5}$,
    I. Kayo$^{4,6}$, R. H. Wechsler$^{7}$\footnote{Hubble Fellow}, F. Marin$^7$, G. Kulkarni$^{2}$, M. Blanton$^{8}$, A. J.
    Connolly$^{9}$, J. P. Gardner$^{10}$, B. Jain$^{3}$, C. J. Miller$^{11}$, A. W.
    Moore$^{5}$, A. Pope$^{12,13}$, J. Pun$^{13,14}$, D. Schneider$^{15}$, J. Schneider$^{5}$,
    A. Szalay$^{12}$, I. Szapudi$^{13}$, I. Zehavi$^{16,17}$, N. A. Bahcall$^{18}$, I. Csabai$^{19}$, J. Brinkmann$^{20}$}
  \vspace*{6pt} \\
  $^{1}$Institute of Cosmology \& Gravitation, University of Portsmouth, Portsmouth, PO1 2EG, UK\\
  $^{2}$Dept. of Physics, Carnegie Mellon University, 5000 Forbes Ave., Pittsburgh, PA15217, USA\\
  $^{3}$Dept. of Physics and Astronomy, University of Pennsylvania, Philadelphia, PA15105, USA\\
  $^{4}$Dept. of Physics, School of Science, University of Tokyo, Tokyo 113-0033, Japan\\
  $^{5}$School of Computer Science, Carnegie Mellon University, 5000 Forbes Ave., Pittsburgh, PA15217, USA\\
  $^{6}$Dept. of Physics and Astrophysics, Nagoya University, Chikusa, Nagoya 464-8602, Japan\\
  $^{7}$Dept. of Astronomy \& Astrophysics, University of Chicago, 5640 S. Ellis Ave., Chicago, IL60637, USA\\
  $^{8}$Center for Cosmology and Particle Physics, Dept. of Physics, New York University, NY10003, USA\\
  $^{9}$Dept. of Physics and Astronomy, University of Pittsburgh, 3941 O'Hara Street, Pittsburgh, PA15260, USA\\
  $^{10}$Pittsburgh Supercomputing Center, 4400 Forbes Ave., Pittsburgh, PA15213, USA\\
  $^{11}$Cerro-Tololo Inter-American Observatory, NOAO, Casilla 603, LaSerena, Chile\\
  $^{12}$Dept. of Physics and Astronomy, John Hopkins University, 400 N. Charles Street, Baltimore, MD21218,USA\\
  $^{13}$Institute for Astronomy, University of Hawaii, 2680 Woodlawn Drive, Honolulu, HI96822, USA\\
  $^{14}$Purple Mountain Observatory, Nanjing 210008, China\\
  $^{15}$Dept. of Astronomy \& Astrophysics, Penn State University,  525 Davey Lab, University Park, PA 16802, USA.\\
  $^{16}$Dept. of Astronomy, Case Western Reserve University, Cleveland, OH44106, USA.\\
  $^{17}$Steward Observatory, University of Arizona, 933 N. Cherry Ave., Tucson AZ85721, USA.\\
  $^{18}$Dept. of Astrophysical Sciences, Princeton University, Peyton Hall, Ivy Lane, Princeton, NJ08544, USA\\
  $^{19}$Department of Physics of Complex Systems, Eotvos University, Pazmany Peter setany 1, H-1518 Budapest\\
  $^{20}$Apache Point Observatory, 2001 Apache Point Road, P.O. Box 59, Sunspot, NM88349, USA\\
  \vspace*{-1.0truecm} }



\maketitle


\begin{abstract}
  We present measurements of the normalised redshift--space
  three--point correlation function ($Q_z$) of galaxies from the Sloan
  Digital Sky Survey (SDSS) main galaxy sample. These measurements
  were possible because of a fast new N--point correlation function
  algorithm (called {\it npt}) based on multi--resolutional k-d trees.
  We have applied {\it npt} to both a volume--limited (36738 galaxies
  with $0.05\le z\le 0.095$ and $-23 \le {M_{^{0.0}r}} \le -20.5$) and
  magnitude--limited sample (134741 galaxies over $0.05 \le z \le
  0.17$ and $\sim M^{*}\pm1.5$) of SDSS galaxies, and find consistent
  results between the two samples, thus confirming the weak luminosity
  dependence of $Q_z$ recently seen by other authors.  We compare our
  results to other $Q_z$ measurements in the literature and find it to
  be consistent within the full jack--knife error estimates.  However,
  we find these errors are significantly increased by the presence of
  the ``Sloan Great Wall'' (at $z\sim0.08$) within these two SDSS
  datasets, which changes the 3--point correlation function (3PCF) by
  70\% on large scales ($s\ge10h^{-1}$ Mpc).  If we exclude this
  supercluster, our observed $Q_z$ is in better agreement with that
  obtained from the 2dFGRS by other authors, thus demonstrating the
  sensitivity of these higher--order correlation functions to
  large--scale structures in the Universe.  This analysis highlights
  that the SDSS datasets used here are not ``fair samples'' of the
  Universe for the estimation of higher--order clustering statistics
  and larger volumes are required.  We study the shape--dependence of
  $Q_z(s,q,\theta)$ as one expects this measurement to depend on scale
  if the large scale structure in the Universe has grown via
  gravitational instability from Gaussian initial conditions. On small
  scales ($s\le6h^{-1}$ Mpc), we see some evidence for
  shape--dependence in $Q_z$, but at present our measurements are consistent with a constant within the errors
  ($Q_z\simeq0.75\pm0.05$). On scales $>10h^{-1}$ Mpc, we see
  considerable shape--dependence in $Q_z$. However, larger samples are
  required to improve the statistical significance of these
  measurements on all scales.
\end{abstract}

\section{Introduction}
Correlation functions are some of the most commonly used statistics in
cosmology.  They have a long history in quantifying the clustering of
galaxies in the Universe (see Peebles 1980). 
There is a hierarchy of correlation functions.  
The two--point correlation function (2PCF) compares the number of 
pairs of data points, as a function of separation, with that expected 
from a Poisson distribution.  
Next in the hierarchy is the 3--point correlation function (3PCF), 
which compares the number of data triplets, as a function of their
triangular--configuration, to that expected from Poisson.  
Higher-order correlations are defined analogously.  

As discussed by many authors, the higher--order correlation functions
contain a variety of important cosmological information, which 
complements that from the 2PCF \citep{GP1977,BS1989}.  
These include tests of Gaussianity and the determination of galaxy 
bias as a function of scale
\citep{Suto1993,JB1998,TJ2003,JB2004,Kayo2004,LS2004}. 
Such tests can also be performed using the Fourier-space equivalent 
of the 3PCF, the bi-spectrum \citep{PEEBLES1980,Sc1999,PSCz,Verde2002} 
or other statistics such as the void probability distribution and 
Minkowski functionals\citep{MBW1994}.  
Recent results from these complementary statistics using the SDSS 
main galaxy sample include \cite{H2002,H2003,H2005} and \cite{Park2005}.

While the 3PCF is easier to correct for survey edge effects than these
other statistics, measurements of the 3PCF have been limited by the
availability of large redshift surveys of galaxies (see Szapudi,
Meiksin \& Nichol 1996, Frieman \& Gaztanaga 1999, Szapudi et al.
2002 for 3PCF analyses of large solid angle catalogues of galaxies) and 
the potentially prohibitive computational time needed to count all
possible triplets of galaxies (naively, this count scales as $O(N^3)$, 
where $N$ is the number of galaxies in the sample).

In this paper, we resolve these two problems through the application
of a new N--point correlation function algorithm \citep{MOORE2001} to
the galaxy data of the Sloan Digital Sky Survey (SDSS; York et al.
2000). We present herein measurements of the 3PCF from the SDSS main 
galaxy sample.  Our measurements illustrate the sensitivity of the 
3PCF to known large-scale structures in the SDSS \citep{Gott2005}.  
They are complementary to the work of \cite{Kayo2004} who explicitly 
explored the luminosity and morphological dependence of the 3PCF 
using SDSS volume--limited galaxy samples.  These measurements of 
the 3PCF will help facilitate constraints on the biasing of galaxies 
and will aid in the development of theoretical predictions for
the higher--order correlation functions \citep{Sc2001,TJ2003}.  
Throughout this paper, we use the dimensionless Hubble constant 
$h \equiv H_{\rm 0}/100\,{\rm km\,s^{-1}\,Mpc^{-1}}$, 
the matter density parameter $\Omega_{\rm m}=0.3$, 
and the dimensionless cosmological constant $\Omega_\Lambda=0.7$, 
unless stated otherwise.

\section{The 3PCF Computational Algorithm}

To facilitate the rapid calculation of the higher--order correlation
functions, we have designed and implemented a new N--point correlation
function (NPCF) algorithm based on k-d trees, which are
multi--dimensional binary search tree for points in a k-dimensional
space.  The k-d tree is composed of a series of inter--connected
nodes, which are created by recursively splitting each node along its
longest dimension, thus creating two smaller child nodes.  This
recursive splitting is stopped when a pre-determined number of data
points is reached in each node (we used $\leq20$ data points herein).
For our NPCF algorithm, we used an enhanced version of the k-d tree
technology, namely multi--resolutional k-d trees with cached
statistics (mrkdtree), which store additional statistical information
about the search tree, and the data points in each node, {\it e.g.},
we store the total count and centroid of all data in each node. 

The key to our NPCF algorithm is to use multiple mrkdtrees together,
and store them in main memory of the computer (rather than on disk),
to represent the required N--point function, {\it e.g.}, we use 3
mrkdtrees to compute the 3PCF, 4 mrkdtrees for the 4PCF, and so on.
The computational efficiency is increased by pruning these trees
wherever possible, and by using the cached statistics on the tree as
much as possible.  The details of mrkdtrees and our NPCF algorithm (known as {\it npt})
have already been outlined in several papers
\citep{MOORE2001,NICHOL2003,GRAY2004}. Similar tree--based
computational algorithms have been discussed by \cite{Szapudi2001}.

\begin{figure*}
\centerline{\psfig{file=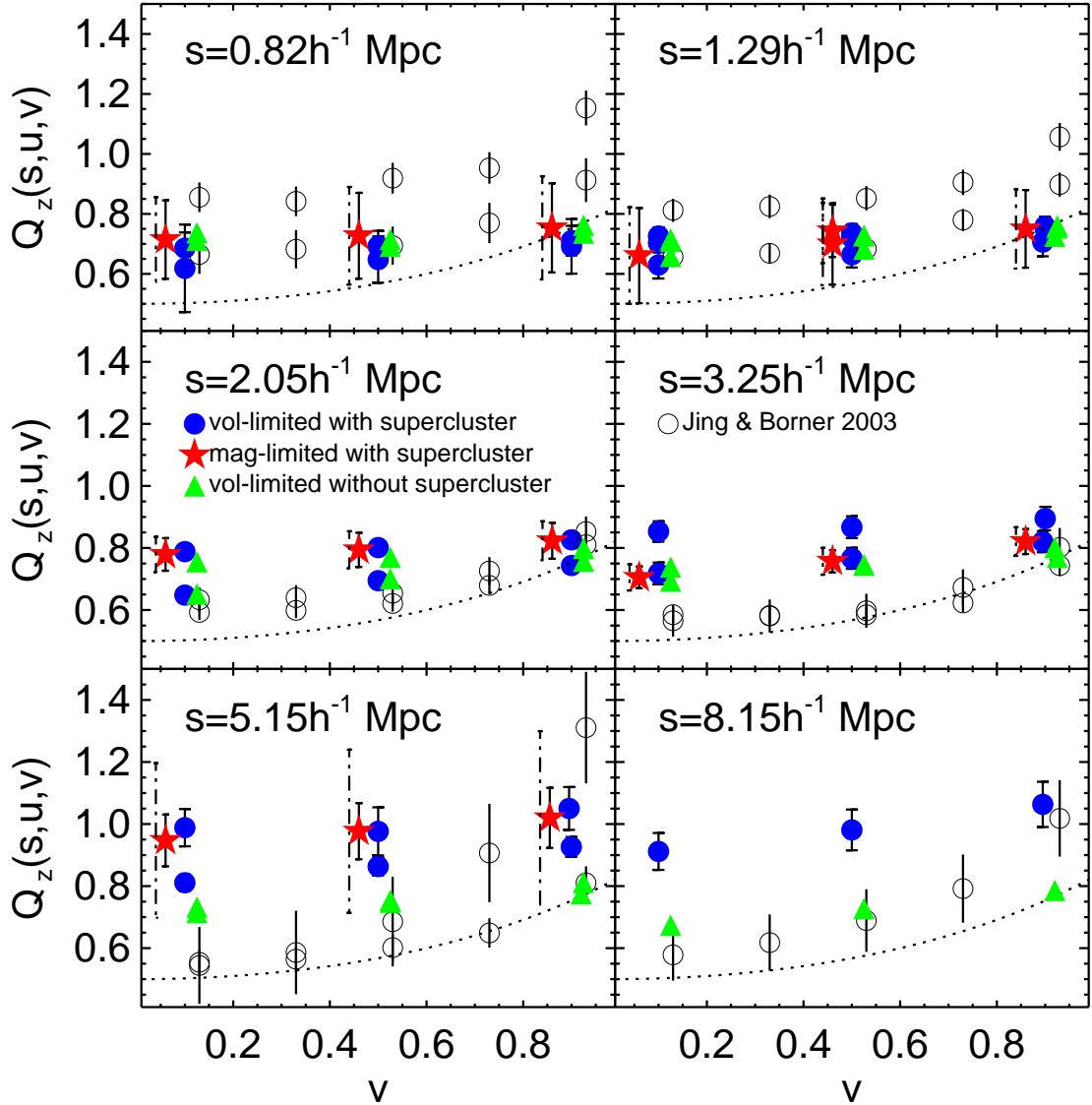}} 
\caption{Our SDSS measurements of the normalised redshift--space 3PCF as a
  function of triangle configuration, {\it i.e.}, $Q_z(s,u,v)$. We
  compare our measurement with that of Jing \& B\"orner (1998, 2004)
  for both the 2dFGRS (open circles and error bars) and LCRS (dashed
  line) galaxy surveys. These two estimates of $Q_z(s,u,v)$ do not
  agree because of the different passbands used for these two surveys.
  We also provide two values of $u$ (namely $u=1.29$ and $u=3.04$)
  from the Jing \& B\"orner (2004) data, {\it i.e.}, we have not
  plotted the $u=2.09$ data to avoid over-crowding.  The binning has
  been chosen to be identical to that of Jing \& B\"orner (2004). The
  solid (blue) circles are the SDSS $Q_z$ for the volume--limited
  sample as discussed in Section \ref{data}, while the solid (red)
  star symbols are the SDSS $Q_z$ for the SDSS magnitude limited
  sample.  The solid error bars shown on these data--points are
  estimated using jack--knife re--sampling (see text), but with
  sub--regions 3 and 4 in Figure \ref{jk} removed ({\it i.e.},
  excluding the supercluster from these error bars).  For comparison,
  the dot--dashed error bars on the red star symbols are our estimate
  of the jack--knife errors (the diagonal elements of the covariance
  matrix) using all 14 sub--regions to estimate the error ({\it i.e.},
  the effect of the supercluster is now including in the size of the
  error bar). In some cases, the error bars are smaller than the
  plotting symbols.  The solid (green) triangle symbols are the SDSS
  $Q_z$ for the jack--knife re--sample excluding sub--regions 3 and 4
  \label{jingplot}}
\end{figure*}

\begin{figure*}
  \centerline{\psfig{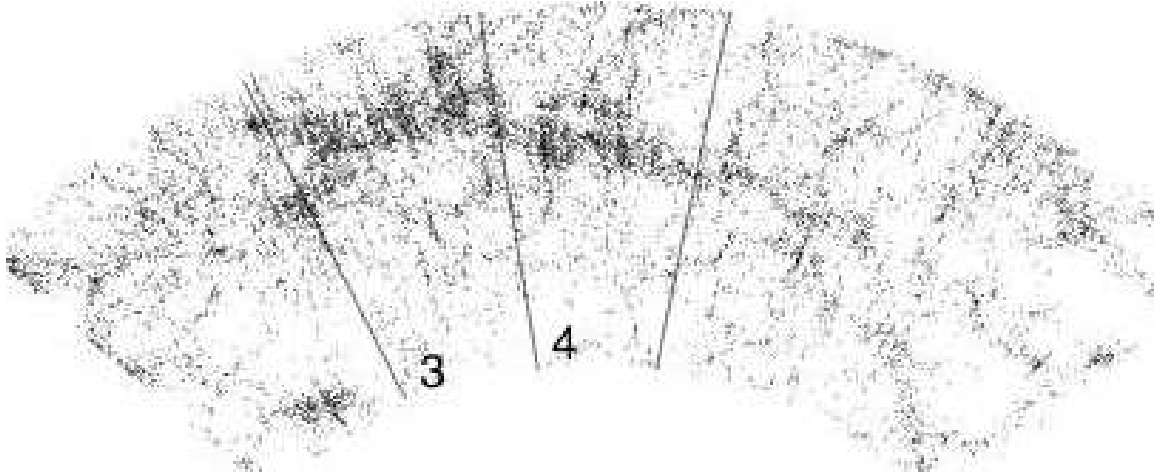}}
\caption{Part of the SDSS volume--limited sample defined in Section \ref{data}. This redshift slice is approximately 500 by 200 $h^{-1}$ Mpc in the dimensions shown, and $\sim100h^{-1}$ Mpc thick (although here we have collapsed the slice in this $3^{rd}$ dimension).  Most noticeable is the supercluster, which has been called the ``Sloan Great Wall'' by Gott et al. (2005) and is 1.37 billion light years long.  This supercluster is a combination of the Leo A  and SCL126 superclusters (Einasto et al. 2001), and is associated with  tens of known Abell clusters of galaxies.  The two regions labelled  3 and 4 are two of the 14 sub--regions used in deriving the  covariance matrices on our correlation functions as shown as error bars in Figure  \ref{jingplot}.}
\label{jk}
\end{figure*}

\begin{figure*}
  \centerline{\psfig{file=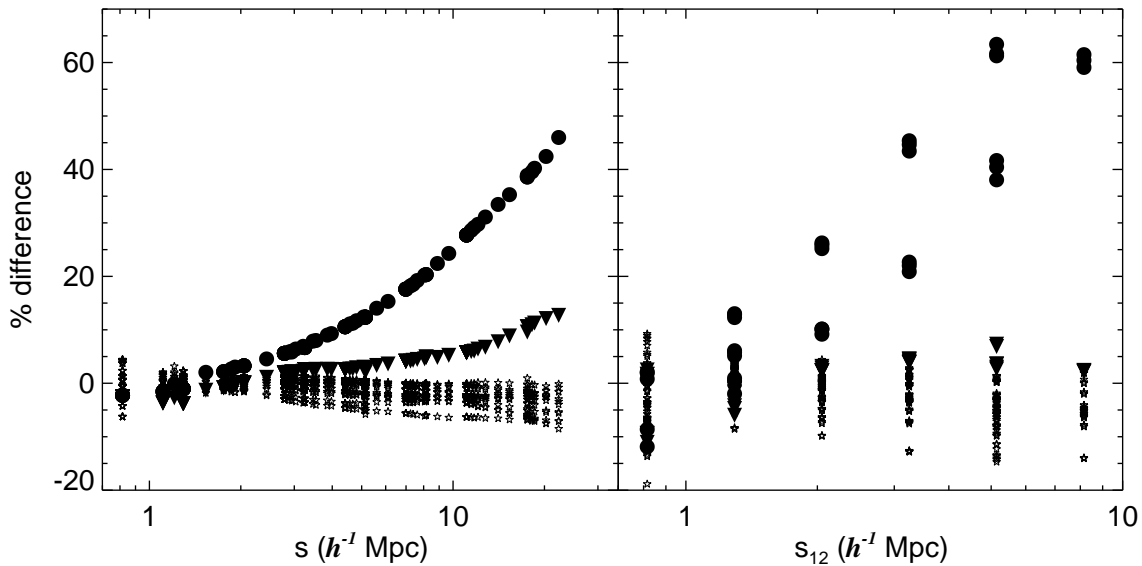,width=6.0in}}
\caption{(Left) The percentage difference between the 2PCFs for the 14
  SDSS jack--knife datasets and the 2PCF as measured for the whole
  dataset (without any sub--regions excluded). The open stars are for
  the 12 jack--knife datasets with the supercluster shown in Figure
  \ref{jk} included, while solid triangles are for the dataset with
  sub--region 4 excluded (Figure \ref{jk}) and solid circles are for
  the dataset with sub--region 3 excluded. (Right) The percentage
  difference between the 3PCFs for the 14 SDSS jack--knife datasets
  and the 3PCF as measured for the whole dataset. The x-axis
  ($S_{12}$) is the redshift--space distance for the shortest side of
  the triangle (see Eqn 2).The open stars are for jack--knife datasets
  with the supercluster shown in Figure \ref{jk} included, while solid
  triangles are for the dataset with sub--region 4 excluded (Figure
  \ref{jk}) and solid circles are for the dataset with sub--region 3
  excluded. All triangle configurations are plotted here, {\it i.e.},
  one point per triangle configuration in Figure 1, which explains why
  there are many data points with the same values of $s$ and
  $s_{12}$.}
\label{plotcorr}
\end{figure*}

\begin{figure*}
  \psfig{file=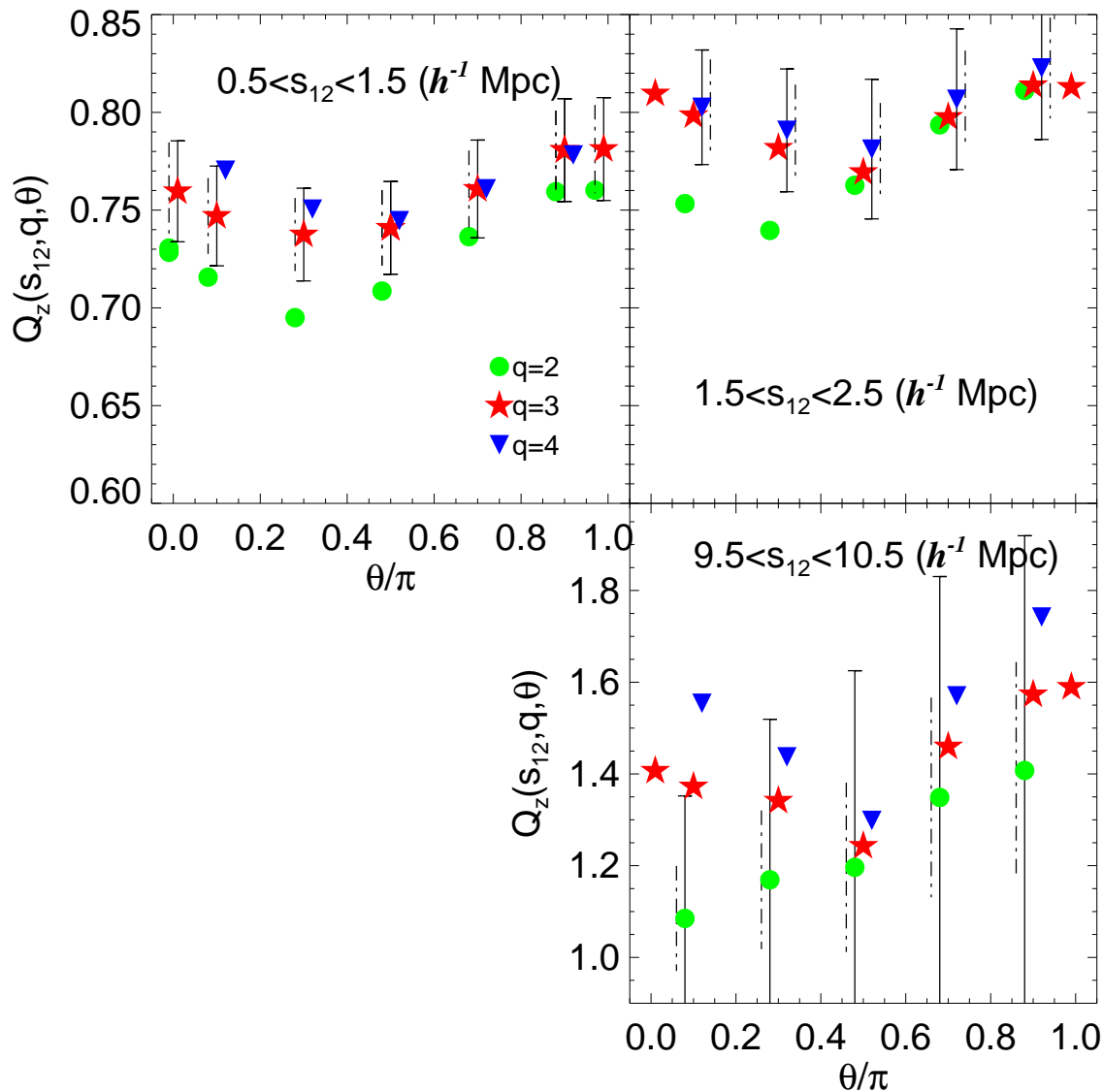}
\caption{The SDSS $Q_z$ as a function of $\theta$, $q$ and $s_{12}$ for the Pope et al. sample discussed in the text. We show 3 bins in $s_{12}$ (as labeled), while  $\theta$ is given in radians along the x-axis (bin width of 
  $\Delta\theta\pm0.05$ radians about the central value plotted). The
  solid (green) circles are for $q=2$, solid (red) stars are for $q=3$
  and the solid (blue) triangles are for $q=4$ ($\Delta q\pm0.5$ about
  the central value). We also include a set of data points at the
  extremes of the $\theta$ range, specifically $0.0<\theta<0.02$ and
  $0.98<\theta<1.00$. The $q=2$ and $q=4$ data points are offset by
  $-0.02$ and $+0.02$ radians respectively to reduce overcrowding.
  Likewise, we only plot the full error bars (solid lines) on a
  fraction of the data points. The dot-dashed error bars have been
  calculated with sub--regions 3 and 4 omitted (see Figure 2). }
\label{tjplot}
\end{figure*}

\begin{figure*}
  \psfig{file=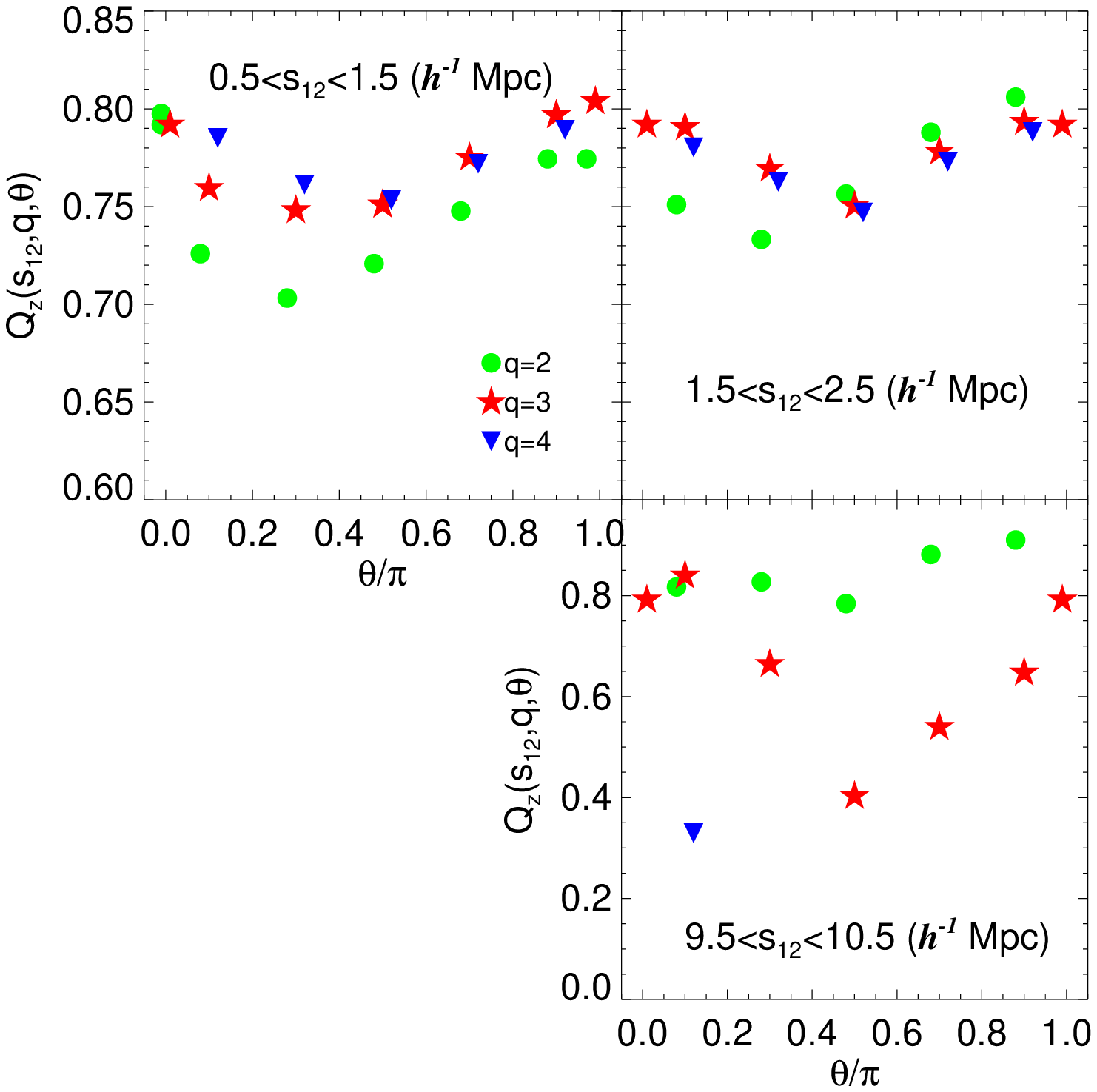} \caption{The same as Figure
    \ref{tjplot} but now with sub--region 3 omitted from the calculation
    of $Q_z(s_{12},q,\theta)$ for the Pope et al. (2004) sample. We
    have not plotted error bars.} \label{tjplotjk4}
\end{figure*}

\section{SDSS Data}
\label{data}

The details of the SDSS survey are given in a series of technical
papers by \cite{fuk96, gunn98, Y2000, hogg01, Strauss2002, smith02,
  pier03, B2003b, ivezic04, DR3}. For the computations discussed
herein, we use two SDSS catalogues. The first is a volume--limited
sample of 36738 galaxies in the redshift range of $0.05 \le z \le
0.095$ and absolute magnitude range of $-23 \le {M_{^{0.0}r}} \le
-20.5$ (for $h=0.7$ and the $z=0.0$ SDSS $r$ filter, or $^{0.0}r$ in
\cite{B2003b} terminology\footnote{\cite{B2003b} use redshifted SDSS
  filters to minimise the effects of k--corrections. As discussed in
  their paper, they propose the use of an SDSS filter set redshifted
  to $z=0.1$ for their ``rest--frame'' quantities. These filters are
  written as $^{0.1}u,^{0.1}g,^{0.1}r,^{0.1}i,^{0.1}z $}), covering
2364 deg$^2$ of the SDSS photometric survey.  All the magnitudes were
reddening corrected using \cite{SFD98}, and the {\tt k-corrected
  v1\_16} software \citep{B2003b}.  The second sample is the same as
``Sample 12'' used by \cite{Pope2004} and contains 134741 galaxies
over 2406 deg$^2$.  This latter sample is not volume--limited, but is
constrained to the absolute magnitude range of $-22 \le {M_{^{0.1}r}}
\le -19$ (or $M^{*}\pm1.5$ magnitudes) for $h=1$, and using the
$z=0.1$ SDSS r filter system, or $^{0.1}r$ \citep{B2003b,Zehavi2005}.
To compare the two samples, our volume--limited sample has the
absolute magnitude range of $-23.54 \le {M_{^{0.1}r}} \le -21.04$ in
the same $^{0.1}r$ filter as used for the Pope et al. sample; assuming
a conversion of ${^{0.1}}r\simeq {^{0.0}}r + 0.23$ for the SDSS main
galaxy sample with a median color at $z=0.0$ of ${^{0.0}}(g-r)\simeq
0.8$. This gives a mean space density of $8.25\times 10^{-3}\,h^{3}$
Mpc$^{-3}$, which is comparable to the space densities of the SDSS
main galaxy sample given in Table 2 of \cite{Zehavi2005}.

We have made no correction for missing galaxies due to
fibre--collisions (i.e., two SDSS fibres can not be placed closer than
55 arcseconds on the sky). We do not expect this observational
constraint to bias our correlation functions as the adaptive tilting
of SDSS spectroscopic plates reduces the problem to $\simeq7\%$ of
possible target galaxies being missed (see Blanton et al.  2003a for
details). Furthermore, this bias will only affect pairs of galaxies
separated by less than 100$h^{-1}$ kpc, which is significantly
smaller than the scales studied herein.  In each case, we also
constructed catalogues of random data points (containing $8\times10^5$
points) over the same area of the sky and with the same selection
function as discussed in \cite{Pope2004}.  These random catalogues are
then used to calculate edge effects on the N--point correlation
functions using the estimators presented in \cite{SS1998}.

\section{Results}


There are two common parametrizations of $Q_z$.  One defines 
\begin{equation}
s=s_{12}, \quad u=\frac{s_{23}}{s_{12}}, \quad{\rm and}
\quad v = \frac{s_{31} - s_{23}}{s_{12}},
\end{equation}
where $s_{12}$, $s_{23}$ and $s_{31}$ are the three sides of a 
triangle in redshift space.  
Then $Q(s,u,v)$ is defined by the ratio of the 3PCF 
$\zeta(s_{12}, s_{23}, s_{31})$, to sums of products of 2PCFs 
(e.g. $\xi(s_{12})\xi(s_{13})$ and permutations):  
\begin{equation}
 Q_z(s,u,v) \equiv \frac{\zeta(s_{12}, s_{23}, s_{31})}
  {\xi(s_{12})\xi(s_{23})+\xi(s_{23})\xi(s_{31})+\xi(s_{31})\xi(s_{12})}.
\label{eq:Qdef}
\end{equation}
The second parametrization has $Q_z(s,q,\theta)$ 
with $s=s_{12}$ being the shortest side of the redshift-space triangle, 
$q=s_{23}/s_{12}$, and $\theta$ the angle between these two sides 
($s_{12}$ and $s_{23}$).  

Figure \ref{jingplot} shows $Q_z(s,u,v)$ for both our volume--limited
sample (filled circles) and the \cite{Pope2004} sample (filled stars). 
Different panels show results for a range of triangle configurations.  
To facilitate a direct comparison with results from the literature, 
we have used the same binning scheme as \cite{JB1998,JB2004}, in their 
analyses of the Las Campanas Redshift Survey (LCRS) and 2dF Galaxy 
Redshift Survey (2dFGRS). The open circles show their results.  
Overall, our $Q_z(s,u,v)$ values are consistent with theirs, but 
with some obvious disagreements.  For example, on large scales 
($s_{12} > 10 h^{-1} {\rm Mpc}$), we find larger $Q_z\sim 1$, 
while \cite{JB2004} find much smaller values.  
Although the different selection passbands of the 2dFGRS ($b_j$) 
and SDSS ($r-$band) might account for this difference, 
it cannot account for the disagreement with the LCRS measurements 
of \cite{JB1998} since the LCRS was also $r$--band selected.

To quantify the disagreement, we estimated the covariances of our 
3PCF estimates using the jack--knife re--sampling technique (discussed 
in detail in \cite{Scranton2002} and \cite{Zehavi2002,Zehavi2005}). 
Briefly, the jack-knife resampling technique provides an estimate of 
the ``cosmic variance'' within a sample.  It is calculated by splitting 
the dataset into sub--regions and then measuring the variance seen 
between the estimated correlation functions as sub--regions are 
omitted one-by-one (therefore, if there are $N$ subregions, there are 
$N$ correlation function estimates).  
As shown in Figure 2 of \cite{Zehavi2005}, the jack-knife errors accurately
reproduce the ``true error'' (the dispersion measured between 100 mock
galaxy catalogues), especially for the diagonal terms of the covariance 
matrix of the 2PCF on large scales, (for $r>0.5h^{-1}$Mpc, the 
difference between the two error estimates is always $<10\%$).  
In what follows, we assume that the jack-knife error estimates are 
also accurate for the 3PCF.  

The SDSS dataset is built-up of thin ``wedge-shaped'' regions that 
are 2.5 degrees thick in declination and hundreds of degrees wide in 
right ascension (see York et al. 2000).  
We divided the total volume of our volume limited catalogue up into 
14 sub--regions when estimating the covariance matrix.  
These were selected in Right Ascension along 
the SDSS scans.  To illustrate, Figure~\ref{jk} shows one of the 
redshift wedges; two of the sub--regions (namely sub--regions 3 
and 4) are highlighted to provide an impression of the typical 
size of a subregion, but also because these two particular regions 
will feature prominently in what follows.  

The error bars shown in Figure \ref{jingplot} show the diagonal 
elements of the covariance matrices we estimate from the jack-knife 
method.  The sizes of these diagonal elements (as well as the 
off--diagonal elements) are extremely sensitive to the inclusion or 
exclusion of sub--regions 3 and 4.  
This sensitivity is quantified in Figure \ref{plotcorr} which 
shows the scatter between the 14 2PCFs and 3PCFs used to
construct the covariance matrices. The scatter in the 2PCFs between 12
of the 14 jack--knife datasets, which contain the supercluster seen in
Figure \ref{jk}, is less than 10\% on all scales probed herein ($s\le
40\,h^{-1}$ Mpc) which is consistent with the findings of
\cite{Zehavi2005}. The 2 datasets which exclude sub--regions 3 and 4,
have significantly different 2PCFs, up to 40\% different on the
largest scales, which is again consistent with \cite{Zehavi2005} who
find that this supercluster greatly affects their 2PCF on large scales
and is not accounted for by their estimates of the jack--knife errors.
The effect on the 3PCF of the ``Sloan Great Wall'' is much greater.  
The jack--knife datasets that exclude sub--regions 3 and 4
(which contain the supercluster) differ by up to 70\% (on large
scales) compared to all other 3PCFs.  

In Figure \ref{jingplot}, we show the normalised 3PCF $Q_z$ for the
whole dataset as well as for the datasets with sub--regions 3 and 4
excluded.  With the bulk of this supercluster excluded, the SDSS 3PCF
has much lower $Q_z(s,u,v)$ values on large scales and is now in good
agreement with the \cite{JB2004} 2dFGRS 3PCF on these large scales.
This is also demonstrated in the error bars shown in Figure
\ref{jingplot} which were estimated using all 14 jack--knife datasets
(dot--dashed error bars) and for the 12 jack--knife datasets (solid
error bars) which excluded the supercluster ({\it i.e.}, sub--regions
3 and 4 removed). As expected, the sizes of these error bars are
sensitive to the inclusion of the supercluster: if we exclude the
supercluster, then our error bars are similar to those of
\cite{JB2004}, who assume an analytical approximation for their
errors.  In addition, \cite{JB2004} used the 100k data release of the
2dFGRS and excluded areas of the 2dFGRS with $R(\theta)<0.1$ (areas
with low redshift completeness). As shown in Figure 15 of
\cite{Colless2001}, the northern strip of the 2dFGRS 100k data release
has a large hole in its coverage between 12.5hrs and 13.5hrs in RA
(due mainly to tilting constraints), which coincides with sub--region
3 in Figure \ref{jk}.  Therefore, the sample used by \cite{JB2004}
does {\em not} include the main core of the ``Sloan Great Wall'' and
explains why our measurements of the 3PCF agree with theirs 3PCF when
we exclude sub--regions 3 \& 4.

\cite{Baugh2004}, \cite{Croton2004}, and \cite{Gaz2005} present an 
analysis of the higher--order correlation functions for the full 
2dFGRS catalogue. In Figure 1 of \cite{Baugh2004}, the 
``Sloan Great Wall'' is visible in the NGP strip of the full 2dFGRS. 
\cite{Baugh2004} also found that the presence of this supercluster, 
and another in the 2dFGRS SGP area, significantly affected their 
measurement of the higher--order correlations on scales $>4\,{h^{-1}}$Mpc, 
consistent with our findings in Figures \ref{jingplot} and \ref{plotcorr} 
(see also \cite{Gaz2005}). 
The influence of these superclusters on the higher--order correlation 
functions indicates that we have not yet reached a ``fair sample'' of 
the Universe with the 2dFGRS and SDSS samples used herein.  This was 
also examined by \cite{H2003} using the Minkowski Functions of the 
SDSS galaxy data (see their Fig.8).


\section{Discussion}


In Figure \ref{jingplot} we find similar $Q_z(s,u,v)$ values for the
two different samples discussed in Section \ref{data}, even though the
Pope et al. sample probes $\sim M^*$ galaxies, while our
volume--limited sample traces more luminous galaxies at $M_{^{0.1}r}
\le -21$.  This confirms the findings of \cite{Kayo2004} and
\cite{JB2004} that there is no strong luminosity--dependence in the
$Q_z(s,u,v)$ parameter (from $-23 \le M_{^{0.1}r} \le -19$).
\cite{Croton2004} also reports a weak luminosity dependence in the
volume--averaged 3PCF, which could be consistent with our measurements
given the error bars (see also \cite{Gaz2005}).  The lack of strong
luminosity dependence in 3PCF may be surprising given the strong
luminosity dependence seen in the 2dFGRS and SDSS 2PCFs
\citep{Norberg2001,Zehavi2005}.  
\cite{Kayo2004} discuss this behaviour further and conclude that
galaxy bias must be complex on weakly non--linear to non--linear 
scales (but see \cite{Verde2002, Croton2004, Gaz2005} for alternative 
interpretations).  We will explore this weaker luminosity dependence 
in future papers.

Figure \ref{tjplot} presents the shape--dependence of $Q_z$ for the 
Pope et al. SDSS sample of galaxies, using the second of the two 
common conventions for $Q_z$.  Recall that this parametrization has 
$Q_z(s,q,\theta)$ with $s=s_{12}$ being the shortest side of the 
redshift-space triangle, $q=s_{23}/s_{12}$ and $\theta$ the angle 
between $s_{23}$ and $s_{12}$.  Our choice of triangles is 
motivated by Figure 5 in the halo-model \citep{CS2002} based 
analysis of \cite{TJ2003}, (although their analysis was restricted 
to real-space rather than redshift-space triangles). To minimize 
overcrowding, we only show a subset of the error bars (the diagonal 
of the covariance matrix) on these data points. We also show the 
same error bars but with sub--regions 3 and 4 omitted from the 
calculation of the covariance matrix.  (Figures \ref{jingplot} 
and \ref{jk} show that these two estimates of the error
are similar on small scales but become significantly different on 
large scales.)

On small scales ($s_{12}<2.5{h^{-1}}$ Mpc), the shape of the
normalised 3PCF is consistent (within the errors) for the different
$q$ values (see Figures \ref{tjplot} and \ref{tjplotjk4}), and is
close to a constant value (within the errors) as a function of
$\theta$, {\it i.e.,} $Q_z(s_{12}<2.5h^{-1} {\rm Mpc}) \simeq
0.75\pm0.05$.  We see some evidence for a ``U-shaped'' behaviour in
$Q_z$ on these small-scales, which is predicted by recent theoretical
models of the 3PCF \citep{GS2005}.  For example, \cite{GS2005} see a
strong ``U--shaped'' pattern in $Q_z$ on small scales, {\it i.e.}, in
Figs 2 \& 3 of their paper, they measure a factor of $\sim2$ increase
in both the $Q_z(\theta\simeq5^{\circ})$ and
$Q_z(\theta\simeq175^{\circ})$ values, relative to the
$Q_z(\theta\simeq 90^{\circ})$ values. We do not see as strong an
effect as they claim, but this could be due to our relatively coarse
binning scheme as \cite{GS2005} claim. We will explore this further in
a future paper, with large datasets from the SDSS, but our $Q_z$ does
have the same qualitative shape as they witnessed.  We also note that
our small--scale $Q_z$ measurements are in excellent agreement with
the 2dFGRS measurements of \cite{Gaz2005}, who also see the same weak
``U--shaped'' behavour (compared to simulations) and also have a near
constant value of $Q_z(s_{12}<6h^{-1} {\rm Mpc}) \simeq 0.75$ for
their two different luminosity bins. This is remarkable agreement
given the differences in the 2dFGRS and SDSS galaxy surveys. Finally,
we comment that our values for $Q_z$ on small--scales are
significantly smaller than the theoretical predictions for $Q$ in
real--space (which are $Q\sim3$), but consistent with the expected
decrease in $Q$ as one moves to redshift--space (see Figure 2 of
\cite{GS2005}). The value and shape of our $Q_z$ measurements are
robust to the omission of the supercluster (see Figure
\ref{tjplotjk4}).

The lack of any strong small--scale shape dependence of $Q_z$ is
consistent with the 2dFGRS findings of \cite{Croton2004} and
\cite{Baugh2004}, using volume--averaged 3PCFs. They found that the
volume--averaged 3PCF scaled as,

\begin{equation}
\xi_3(s)\simeq S_3\,\xi_2(s)^2,
\label{scaling}
\end{equation}

\noindent where $S_3$ displayed an weak luminosity dependence. Assuming 
little shape--dependence in $Q_z(s,q,\theta)$, then we can relate $S_3$ 
to $Q_z$ by assuming the denominator in Eqn 1 of $Q_z$ simply becomes
$\sim 3\langle\xi_2(s)\rangle^2$, and thus $S_3\simeq3\,Q_z$.  The
value of $S_3=1.95\pm0.18$ derived for $L^{\star}$ galaxies in the
2dFGRS volume--averaged 3PCF \citep{Baugh2004} is therefore in good
agreement (within the errors) with our measured value of
$Q_z\simeq0.75$ on small scales for the Pope et al.  sample (Figure
\ref{tjplot}), which was designed to probe $\sim L^*$ in the SDSS.
This again demonstrates the relative insensitivity of the 3PCF (in
redshift--space) to the details of the selection of the galaxy sample.
The simple scaling relationship given in Eqn \ref{scaling} is expected
for hierarchical structure formation models originating from Gaussian
initial conditions \citep{PEEBLES1980, Baugh2004}.

On larger scales (10 $h^{-1}$ Mpc), the amplitude and
shape--dependence of $Q_z$ changes significantly once the supercluster
has been removed (comparing Figures \ref{tjplot} and \ref{tjplotjk4}).
For example, for the $q=2$ triangle configurations (circle symbols),
the ``U--shape'' in $Q_z$ is only seen once the core of the ``Sloan
Great Wall'' has been removed. Likewise, ``U--shape'' behavour of
$Q_z$ for the $q=3$ triangle configurations (star symbols) is enhanced
(by nearly a factor of 3) when the supercluster is removed, and is
then in better agreement with the numerical simulations of
\cite{GS2005} and measurements for the 2dFGRS \citep{Gaz2005}.
Therefore, the expected ``U--shaped'' signal in $Q_z$
due to filamentary stuctures in the Universe has been overwhelmed by
the presence of the supercluster, and is only seen when the ``Sloan
Great Wall'' is removed. This indicates that the ``Sloan Great Wall''
has a different topology than filaments ({\it e.g.} sheet--like) or
this difference is caused by the orientation of this supercluster in
the SDSS (it appears to be perpendicular to the line--of--sight).
Overall, the 3PCF is hard to measure on these large scales using the
samples presented herein, and the errors are dominated by the ``Sloan
Great Wall''. Larger samples, in both volume and numbers of galaxies,
are required to explore the shape--dependence of the 3PCF in greater
detail on these large scales, and that should be possible with future
SDSS samples.

\section*{Acknowledgments}

We thank an anonymous referee for their careful reading of the paper
and useful comments.  We thank Y.P. Jing for extensive discussions of
his work and providing his data points. We also thank Carlton Baugh,
John Lacey, Robert Crittenden and Shaun Cole for helpful comments and
discussions about this work. We thank Quentin Mercer III, Rupert Croft
and Albert Wong for their help and assistance in building and running
the astrophysics Beowulf cluster at Carnegie Mellon University which
was used to compute the SDSS 3PCF. We also thank Stuart Rankin and
Victor Travieso for their assistance in running the NPT code on the UK
COSMOS Supercomputer.

RN thanks the EU Marue Curie program for partial funding during this
work. The work presented here was also partly funded by NSF ITR Grant
0121671. RKS was supported in part by NSF grant AST-0520647.  
YS was supported in part by Grants-in-Aid for Scientific Research
from the Japan Society for Promotion of Science (Nos.14102004 and
16340053).
IK acknowledges the support from the Ministry of Education, 
Culture, Sports, Science, and Technology, Grant-in-Aid for 
Encouragement of Young Scientists (No. 15740151).  
RHW is supported by NASA through Hubble Fellowship grant
HST-HF-01168.01-A awarded by the Space Telescope Science Institute.

Funding for the creation and distribution of the SDSS Archive has been
provided by the Alfred P. Sloan Foundation, the Participating
Institutions, the National Aeronautics and Space Administration, the
National Science Foundation, the U.S. Department of Energy, the
Japanese Monbukagakusho, and the Max Planck Society. The SDSS Web site
is http://www.sdss.org/.

The SDSS is managed by the Astrophysical Research Consortium (ARC) for
the Participating Institutions. The Participating Institutions are The
University of Chicago, Fermilab, the Institute for Advanced Study, the
Japan Participation Group, The Johns Hopkins University, the Korean
Scientist Group, Los Alamos National Laboratory, the
Max-Planck-Institute for Astronomy (MPIA), the Max-Planck-Institute
for Astrophysics (MPA), New Mexico State University, University of
Pittsburgh, University of Portsmouth, Princeton University, the United
States Naval Observatory, and the University of Washington.

\section{Appendix A: The 3PCF Data}

We present here the data points from Figures 4 \& 5. We present the
upper and lower limits of the bins used. We stress that these data are
affected by large scale structures in the data and, therefore, should
be used with caution. We present these data to aid in the comparison
with other observations and theoretical predictions.

\begin{table*}
\centering
\caption{The data presented in Figure 4 of this paper}
\begin{tabular}{@{}cccccccc@{}} \hline
$s_{12}^{low}$ & $s_{12}^{high}$ & $q^{low}$ & $q^{high}$ & $\theta^{low}$ & $\theta^{high}$ & $Q_z$($s_{12}$,$q$,$\theta$) & $\delta Q_z$($s_{12}$,$q$,$\theta$) \\ \hline
     0.5 &       1.50 &       1.50 &       2.50 &       0.000 &     0.02 &      0.7284 &     0.0220 \\ 
     0.5 &       1.50 &       1.50 &       2.50 &     0.05 &      0.150 &      0.7157 &      0.1302 \\ 
     0.5 &       1.50 &       1.50 &       2.50 &      0.250 &      0.350 &      0.695 &     0.0226 \\ 
     0.5 &       1.50 &       1.50 &       2.50 &      0.450 &      0.550 &      0.7086 &      0.1033 \\ 
     0.5 &       1.50 &       1.50 &       2.50 &      0.650 &      0.750 &      0.7364 &     0.0244 \\ 
     0.5 &       1.50 &       1.50 &       2.50 &      0.850 &      0.950 &      0.7594 &      0.1024 \\ 
     0.5 &       1.50 &       1.50 &       2.50 &      0.980 &       1.000 &      0.7602 &     0.0258 \\ 
     0.5 &       1.50 &       2.50 &       3.50 &       0.000 &     0.02 &      0.7596 &     0.0258 \\ 
     0.5 &       1.50 &       2.50 &       3.50 &     0.05 &      0.150 &      0.747 &     0.0255 \\ 
     0.5 &       1.50 &       2.50 &       3.50 &      0.250 &      0.350 &      0.7375 &     0.0237 \\ 
     0.5 &       1.50 &       2.50 &       3.50 &      0.450 &      0.550 &      0.7409 &     0.0237 \\ 
     0.5 &       1.50 &       2.50 &       3.50 &      0.650 &      0.750 &      0.7608 &     0.0250 \\ 
     0.5 &       1.50 &       2.50 &       3.50 &      0.850 &      0.950 &      0.7807 &     0.0263 \\ 
     0.5 &       1.50 &       2.50 &       3.50 &      0.980 &       1.000 &      0.7812 &     0.0263 \\ 
     0.5 &       1.50 &       3.50 &       4.50 &     0.050 &      0.150 &      0.7704 &     0.0251 \\ 
     0.5 &       1.50 &       3.50 &       4.50 &      0.250 &      0.350 &      0.7505 &     0.0245 \\ 
     0.5 &       1.50 &       3.50 &       4.50 &      0.450 &      0.550 &      0.7446 &     0.0239 \\ 
     0.5 &       1.50 &       3.50 &       4.50 &      0.650 &      0.750 &      0.7609 &     0.0249 \\ 
     0.5 &       1.50 &       3.50 &       4.50 &      0.850 &      0.950 &      0.7782 &     0.0259 \\ 
      1.50 &       2.50 &       1.50 &       2.50 &     0.05 &      0.150 &      0.7533 &     0.0231 \\ 
      1.50 &       2.50 &       1.50 &       2.50 &      0.250 &      0.350 &      0.7396 &     0.0206 \\ 
      1.50 &       2.50 &       1.50 &       2.50 &      0.450 &      0.550 &      0.7627 &     0.0217 \\ 
      1.50 &       2.50 &       1.50 &       2.50 &      0.650 &      0.750 &      0.7936 &     0.0230 \\ 
      1.50 &       2.50 &       1.50 &       2.50 &      0.850 &      0.950 &      0.8112 &     0.0239 \\ 
      1.50 &       2.50 &       2.50 &       3.50 &       0.000 &     0.02 &      0.8096 &     0.0238 \\ 
      1.50 &       2.50 &       2.50 &       3.50 &     0.050 &      0.150 &      0.7985 &     0.0236 \\ 
      1.50 &       2.50 &       2.50 &       3.50 &      0.250 &      0.350 &      0.7818 &     0.0418 \\ 
      1.50 &       2.50 &       2.50 &       3.50 &      0.450 &      0.550 &      0.7694 &     0.0271 \\ 
      1.50 &       2.50 &       2.50 &       3.50 &      0.650 &      0.750 &      0.7976 &     0.0278 \\ 
      1.50 &       2.50 &       2.50 &       3.50 &      0.850 &      0.950 &      0.8138 &     0.0286 \\ 
      1.50 &       2.50 &       2.50 &       3.50 &      0.980 &       1.00 &      0.813 &     0.0289 \\ 
      1.50 &       2.50 &       3.50 &       4.50 &     0.050 &      0.150 &      0.8026 &     0.0293 \\ 
      1.50 &       2.50 &       3.50 &       4.50 &      0.250 &      0.350 &      0.7908 &     0.0314 \\ 
      1.50 &       2.50 &       3.50 &       4.50 &      0.450 &      0.550 &      0.7812 &     0.0357 \\ 
      1.50 &       2.50 &       3.50 &       4.50 &      0.650 &      0.750 &      0.8067 &     0.0360 \\ 
      1.50 &       2.50 &       3.50 &       4.50 &      0.850 &      0.950 &      0.8227 &     0.0366 \\ 
      9.50 &       10.5 &       1.50 &       2.50 &     0.050 &      0.150 &       1.085 &      0.267 \\ 
      9.50 &       10.5 &       1.50 &       2.50 &      0.250 &      0.350 &       1.16920 &      0.3501 \\ 
      9.50 &       10.5 &       1.50 &       2.50 &      0.450 &      0.550 &       1.19640 &      0.4286 \\ 
      9.50 &       10.5 &       1.50 &       2.50 &      0.650 &      0.750 &       1.34880 &      0.4817 \\ 
      9.50 &       10.5 &       1.50 &       2.50 &      0.850 &      0.950 &       1.40760 &      0.5118 \\ 
      9.50 &       10.5 &       2.50 &       3.50 &       0.000 &     0.02 &       1.40680 &      0.6107 \\ 
      9.50 &       10.5 &       2.50 &       3.50 &     0.050 &      0.150 &       1.37250 &      0.5392 \\ 
      9.50 &       10.5 &       2.50 &       3.50 &      0.250 &      0.350 &       1.34150 &      0.6721 \\ 
      9.50 &       10.5 &       2.50 &       3.50 &      0.450 &      0.550 &       1.24310 &      0.8343 \\ 
      9.50 &       10.5 &       2.50 &       3.50 &      0.650 &      0.750 &       1.45970 &      0.9198 \\ 
      9.50 &       10.5 &       2.50 &       3.50 &      0.850 &      0.950 &       1.57340 &      0.9453 \\ 
      9.50 &       10.5 &       2.50 &       3.50 &      0.980 &       1.00 &       1.58970 &       1.18550 \\ 
      9.50 &       10.5 &       3.50 &       4.50 &     0.050 &      0.150 &       1.55410 &       1.23750 \\ 
      9.50 &       10.5 &       3.50 &       4.50 &      0.250 &      0.350 &       1.43730 &       1.20710 \\ 
      9.50 &       10.5 &       3.50 &       4.50 &      0.450 &      0.550 &       1.29870 &       1.20890 \\ 
      9.50 &       10.5 &       3.50 &       4.50 &      0.650 &      0.750 &       1.57040 &       1.22310 \\ 
      9.50 &       10.5 &       3.50 &       4.50 &      0.850 &      0.950 &       1.74220 &       1.24910 \\ \hline
\end{tabular}
\end{table*}

\begin{table*}
\centering
\caption{The data presented in Figure 5 of this paper}
\begin{tabular}{@{}cccccccc@{}} \hline
$s_{12}^{low}$ & $s_{12}^{high}$ & $q^{low}$ & $q^{high}$ & $\theta^{low}$ & $\theta^{high}$ & $Q_z$($s_{12}$,$q$,$\theta$) & $\delta Q_z$($s_{12}$,$q$,$\theta$) \\ \hline
     0.5 &       1.50 &       1.50 &       2.50 &     0.05 &      0.150 &      0.7259 &     0.0269 \\ 
     0.5 &       1.50 &       1.50 &       2.50 &      0.250 &      0.350 &      0.7032 &     0.0178 \\ 
     0.5 &       1.50 &       1.50 &       2.50 &      0.450 &      0.550 &      0.7208 &     0.0965 \\ 
     0.5 &       1.50 &       1.50 &       2.50 &      0.650 &      0.750 &      0.7477 &     0.0180 \\ 
     0.5 &       1.50 &       1.50 &       2.50 &      0.850 &      0.950 &      0.7744 &     0.0957 \\ 
     0.5 &       1.50 &       2.50 &       3.50 &     0.05 &      0.150 &      0.7595 &     0.0194 \\ 
     0.5 &       1.50 &       2.50 &       3.50 &      0.250 &      0.350 &      0.7481 &     0.0187 \\ 
     0.5 &       1.50 &       2.50 &       3.50 &      0.450 &      0.550 &      0.751 &     0.0193 \\ 
     0.5 &       1.50 &       2.50 &       3.50 &      0.650 &      0.750 &      0.7751 &     0.0197 \\ 
     0.5 &       1.50 &       2.50 &       3.50 &      0.850 &      0.950 &      0.7968 &     0.0202 \\ 
     0.5 &       1.50 &       3.50 &       4.50 &     0.05 &      0.150 &      0.7851 &     0.0199 \\ 
     0.5 &       1.50 &       3.50 &       4.50 &      0.250 &      0.350 &      0.7612 &     0.0209 \\ 
     0.5 &       1.50 &       3.50 &       4.50 &      0.450 &      0.550 &      0.7534 &     0.0213 \\ 
     0.5 &       1.50 &       3.50 &       4.50 &      0.650 &      0.750 &      0.7719 &     0.0221 \\ 
     0.5 &       1.50 &       3.50 &       4.50 &      0.850 &      0.950 &      0.7893 &     0.0226 \\ 
      1.50 &       2.50 &       1.50 &       2.50 &     0.05 &      0.150 &      0.751 &     0.0190 \\ 
      1.50 &       2.50 &       1.50 &       2.50 &      0.250 &      0.350 &      0.7332 &     0.0193 \\ 
      1.50 &       2.50 &       1.50 &       2.50 &      0.450 &      0.550 &      0.7564 &     0.0205 \\ 
      1.50 &       2.50 &       1.50 &       2.50 &      0.650 &      0.750 &      0.788 &     0.0217 \\ 
      1.50 &       2.50 &       1.50 &       2.50 &      0.850 &      0.950 &      0.806 &     0.0225 \\ 
      1.50 &       2.50 &       2.50 &       3.50 &     0.05 &      0.150 &      0.7906 &     0.0224 \\ 
      1.50 &       2.50 &       2.50 &       3.50 &      0.250 &      0.350 &      0.7694 &     0.0356 \\ 
      1.50 &       2.50 &       2.50 &       3.50 &      0.450 &      0.550 &      0.7504 &     0.0230 \\ 
      1.50 &       2.50 &       2.50 &       3.50 &      0.650 &      0.750 &      0.7781 &     0.0242 \\ 
      1.50 &       2.50 &       2.50 &       3.50 &      0.850 &      0.950 &      0.7933 &     0.0249 \\ 
      1.50 &       2.50 &       3.50 &       4.50 &     0.05 &      0.150 &      0.7803 &     0.0245 \\ 
      1.50 &       2.50 &       3.50 &       4.50 &      0.250 &      0.350 &      0.7627 &     0.0234 \\ 
      1.50 &       2.50 &       3.50 &       4.50 &      0.450 &      0.550 &      0.747 &     0.0236 \\ 
      1.50 &       2.50 &       3.50 &       4.50 &      0.650 &      0.750 &      0.773 &     0.0249 \\ 
      1.50 &       2.50 &       3.50 &       4.50 &      0.850 &      0.950 &      0.7882 &     0.0258 \\ 
      9.50 &       10.5 &       1.50 &       2.50 &     0.05 &      0.150 &      0.8173 &      0.1145 \\ 
      9.50 &       10.5 &       1.50 &       2.50 &      0.250 &      0.350 &      0.8273 &      0.1512 \\ 
      9.50 &       10.5 &       1.50 &       2.50 &      0.450 &      0.550 &      0.7843 &      0.1846 \\ 
      9.50 &       10.5 &       1.50 &       2.50 &      0.650 &      0.750 &      0.8816 &      0.2177 \\ 
      9.50 &       10.5 &       1.50 &       2.50 &      0.850 &      0.950 &      0.9102 &      0.2364 \\ 
      9.50 &       10.5 &       2.50 &       3.50 &     0.05 &      0.150 &      0.8398 &      0.2343 \\ 
      9.50 &       10.5 &       2.50 &       3.50 &      0.250 &      0.350 &      0.6645 &      0.2523 \\ 
      9.50 &       10.5 &       2.50 &       3.50 &      0.450 &      0.550 &      0.4031 &      0.299 \\ 
      9.50 &       10.5 &       2.50 &       3.50 &      0.650 &      0.750 &      0.5397 &      0.3653 \\ 
      9.50 &       10.5 &       2.50 &       3.50 &      0.850 &      0.950 &      0.6475 &      0.4195 \\ 
      9.50 &       10.5 &       3.50 &       4.50 &     0.05 &      0.150 &      0.3295 &      0.3391 \\ 
      9.50 &       10.5 &       3.50 &       4.50 &      0.250 &      0.350 &     -0.1497 &      0.4041 \\ 
      9.50 &       10.5 &       3.50 &       4.50 &      0.450 &      0.550 &     -0.8885 &      0.5331 \\ 
      9.50 &       10.5 &       3.50 &       4.50 &      0.650 &      0.750 &      -1.26670 &      0.7041 \\ 
      9.50 &       10.5 &       3.50 &       4.50 &      0.850 &      0.950 &      -1.53840 &      0.8304 \\ \hline
\end{tabular}
\end{table*}

\label{lastpage}
\end{document}